\documentclass[12pt,a4paper]{iopart}

\usepackage{iopams}
\usepackage{setstack}
\usepackage{graphicx}
\usepackage{multicol}
\usepackage{color}
\usepackage{hyperref}

\usepackage{tikz}
\usetikzlibrary{decorations.markings}
\usetikzlibrary{decorations.pathreplacing}

\newcommand{\sgn}{\mathrm{sgn}}
\newcommand{\dd}[1]{\mathrm{d}#1}

\newcommand{\Li}{\mathrm{Li}}
\renewcommand{\Re}{\mathrm{Re}~}
\renewcommand{\Im}{\mathrm{Im}~}
\renewcommand{\[}{[\![}

\newcommand{\tfrac}[2]{\mbox{\small$\frac{#1}{#2}$}}
\renewcommand{\text}[1]{\mathrm{#1}}

\begin{document}
\title{Riemann surface crossover for the spectral gaps of open TASEP}
%\titlerunning{}
\author{Ulysse Godreau, Sylvain Prolhac}
\address{Laboratoire de Physique Th\'eorique, IRSAMC, UPS, Universit\'e de Toulouse, France}
\date{}
%\maketitle
\begin{abstract}
We consider the totally asymmetric simple exclusion process with open boundaries, at the edge of the maximal current phase. Using analytic continuations from the known stationary eigenvalue, we obtain exact expressions for the spectral gaps in the limit of large system size. The underlying Riemann surface, generated by modified Lambert functions, interpolates between the one for periodic TASEP and the one for open TASEP in the maximal current phase.
\end{abstract}

%\today

\begin{section}{Introduction}
The totally asymmetric simple exclusion process (TASEP) \cite{D1998.1,S2001.1,GM2006.1,CMZ2011.1} is a Markov process on a one-dimensional lattice where particles with hard-core interactions randomly hop from any site of the lattice to the next. The simplicity of its definition and the richness of its large scale behaviour makes it a paradigmatic model for non-equilibrium $N$-body stochastic systems.

In this paper we consider TASEP on a finite lattice of $L$ sites with open boundaries, where a particle is injected in the system at site $1$ and removed from the system at site $L$ with respective probability per unit time $\alpha$ and $\beta$. In the $L \rightarrow \infty$ limit,  this process exhibits three different phases \cite{DEHP1993.1} depending on the boundary parameters $\alpha$ and $\beta$, characterized by distinct values for the stationary current and density of particles.

Of special interest is the connection of the exclusion process with the KPZ universality class \cite{KPZ1986.1,HHZ1995.1,KK2010.1,HHT2015.1,QS2015.1,S2019.1}. Indeed, the exclusion process with both forward and backward hopping with respective rates $1$ and $q$ is known \cite{BG1997.1,DM1997.1} to be described by the KPZ equation in the weakly asymmetric scaling $\log q\sim L^{-1/2}$. For fixed $q<1$, and in particular for the totally asymmetric model $q=0$, the model furthermore converges to the KPZ fixed point. Boundary rates $\alpha$ and $\beta$ distant of order $L^{-1/2}$ from their values at the triple point of the phase diagram then correspond \cite{CS2018.1,P2019.1} to fixed, finite values of the slope of the KPZ height at the boundaries. The one-dimensional KPZ dynamical exponent $z=3/2$ appears in particular for the relaxation of KPZ fluctuation to their stationary state, with spectral gaps scaling as $E_n \simeq e_n L^{-3/2}$.

The main goal of this paper is to compute the eigenvalues of the TASEP Markov matrix contributing to the KPZ universal regime at the edge of the maximal current phase. Starting from the expression obtained by Lazarescu and Mallick in \cite{LM2011.1} for the cumulant generating function of the particle current (i.e. the lowest eigenvalue), we obtain by analytic continuations the whole spectrum. Comparison with extrapolated Bethe ansatz numerics confirm that the various branches obtained by analytic continuation are indeed in one to one correspondence with the spectral gaps in the KPZ regime.

The analytic continuation approach has already proved quite successful for computing the spectral gaps of the periodic TASEP \cite{P2020.1,P2020.2} as well as those of the open TASEP in the maximal current phase \cite{GP2020.1}. Analytic continuation between spectral gaps has also been pointed out before in relation with thermodynamic Bethe ansatz \cite{DT1996.1,VT2016.1}. Additionally, branch points for the analytic continuations, where some eigenvalues coincide, correspond to the much studied concept of \emph{exceptional points} encountered generically in the context of non-Hermitian systems, see e.g. \cite{H2012.1}.

It was shown recently by one of the authors in \cite{P2020.1} that KPZ fluctuations with periodic boundaries can be expressed in a unified way in terms of contour integrals on a Riemann surface associated to the analytic continuation of the polylogarithm function $\Li_s$ with half-integer $s$. A motivation for the present work came from the observation \cite{GLMV2012.1,L2015.1} that on the transition lines between the maximal current phase and the high/low density phases, the stationary large deviation function of the current for \emph{open} TASEP is the same as the one for \emph{periodic} TASEP. A natural expectation is then that spectral gaps in the vicinity of the transition lines should be associated to a family of Riemann surfaces interpolating between the Riemann surface defined in \cite{P2020.1} for the periodic case and the Riemann surface constructed in \cite{GP2020.1} for the open case in the maximal current phase. We show that this is indeed the case.

The paper is organized as follows. In section~\ref{sec_large_deviations}, we define the totally asymmetric exclusion process, recall some facts about large deviations of the current, and state our main results. In section~\ref{sec_analytic_continuation}, we construct the analytic continuation of the asymptotic cumulant generating function of the stationary current at the vicinity of the transition line  and construct the associated Riemann surface, leading to conjectural expressions for the spectral gaps. These expressions are finally checked against numerical solutions of the Bethe ansatz equations in section~\ref{sec_numerics}.
\end{section}

\begin{section}{Large deviations in the stationary state}\label{sec_large_deviations}
In this section, after recalling some known facts about the asymmetric exclusion process, we detail the large deviation results obtained by Lazarescu and Mallick in \cite{LM2011.1} for the open TASEP in finite size, and compute their asymptotics for large system size at the edge of the maximal current phase. We then state our main results for the spectral gaps obtained by analytic continuations.

\begin{subsection}{Open TASEP and deformed Markov matrix}
 
 \begin{figure}
  \begin{center}
 \includegraphics[scale=1]{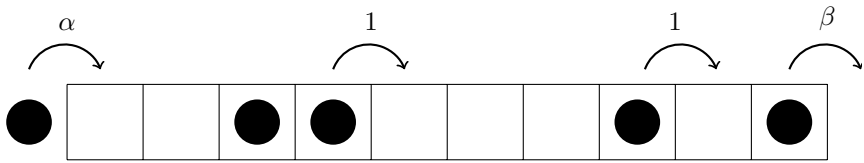}
 \caption{Schematic representation of the dynamical rules of the open TASEP.}
   \label{schema_TASEP}
  \end{center}
\end{figure}

TASEP with open boundaries is a Markov process defined on a one-dimensional lattice of length $L$, in contact at both end with particle reservoirs. The time evolution of the process is defined as follow: during a small time interval $\dd{t}$, a particle at any site $i$, $1 \leq i < L$ may hop to the site $i+1$ if it is empty at rate $1$ (i.e. with probability $\dd{t}$). Moreover, a particle may enter the system at site $1$ with rate $\alpha$ if it is empty, a particle occupying site $L$ may exit the chain with rate $\beta$. The dynamics is summarized in figure~\ref{schema_TASEP}.

At large $L$, the system is either in the low density phase, the high density phase, or the maximal current phase depending on the values of the boundary rates $\alpha$ and $\beta$. The three phases are characterized by different stationary values of the mean current of particle and mean density, see figure \ref{phase_diagram}. The focus of this paper is the transition line between the maximal current phase and the low density phase, which is equivalent to the transition line between the maximal current phase and the high density phase by symmetry between particles and holes. More precisely, we study the crossover regime where the boundary rates scale as
\begin{equation}
\alpha =\frac{1}{2}+\sqrt{\frac{A}{4L}}
\;,\qquad
\beta=\frac{1}{2}+\sqrt{\frac{B}{4L}}
\label{scaling_AB}
\end{equation}
in the limit $B \rightarrow \infty$. It was recently shown \cite{CS2018.1,P2019.1} that with such scaling of the boundary rates, the KPZ height function $h(x,t)$ associated to the asymmetric exclusion process with bulk rates $1$ and $q$ in the weakly asymmetric regime $\log q\sim L^{-1/2}$ is solution to the KPZ equation on the interval $[0,1]$ with Neumann boundary conditions $\partial_x h(x=1,t)=-\infty$ and $\partial_x h(x=0,t)$ an affine function of $\sqrt{A}$ going to $+\infty$ when $A\to\infty$. Note that we are using for convenience a different definition for the coefficients $A$ and $B$ than the one used in \cite{CS2018.1,P2019.1}.

\begin{figure}
  \begin{center}
 \includegraphics[scale=1]{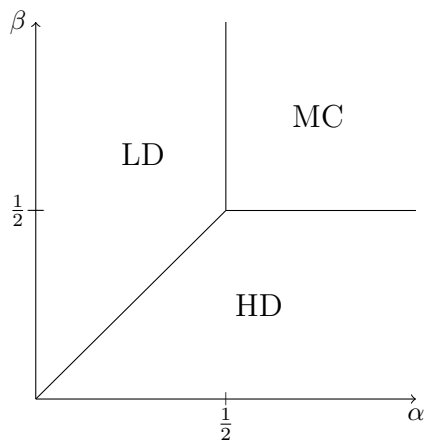}
  \caption{Phase diagram of the open TASEP. In the maximal current phase (MC) $\rho=1/2$, $J=1/4$, in the low density phase (LD) $\rho=1-\beta$, $J=\beta (1-\beta)$ and the high density phase (HD) $\rho=\alpha$, $J=\alpha (1-\alpha)$. The HD and LD phases are separated by a first order transition at the coexistence line $\alpha=\beta<1/2$, while the MC phase is separated from the HD and LD phases by second order transitions.}
  \label{phase_diagram}
  \end{center}
 \end{figure}
 
The state of each site of the system can be described as a vector in $\mathbb{C}^2$ with base vectors $|0 \rangle$ and $|1 \rangle$ representing respectively an empty site and an occupied site. The state of the full system is then described by a vector in the space $(\mathbb{C}^2)^{\otimes L}$. The probability for the system at time $t$ to be in a configuration $\mathcal{C}$ satisfies the master equation
\begin{equation}
 \frac{d}{dt} P_t (\mathcal{C}) = \sum_{\mathcal{C}'} M_{\mathcal{C}, \mathcal{C}'} P_t (\mathcal{C}') \;,
\end{equation}
where $M$ is the Markov matrix of the process, acting on the space $\left( \mathbb{C}^2 \right)^{\otimes L}$, and the summation is over all $2^{L}$ configurations of the model. The Markov matrix can be written as
\begin{equation}
 M = C_1 + \sum_{k=1}^{L-1} W_{k,k+1} + D_L \;.
\end{equation}
The subscript indicate on which site the local operators act non trivially. The matrices above have the following expressions in the local basis $(|0 \rangle ,|1 \rangle )$:
\begin{equation}
\fl\hspace{15mm}
C_1=\left( \begin{array}{cc}
 -\alpha & 0 \\
 \alpha & 0
\end{array} \right)
\qquad
W=\left( \begin{array}{cccc}
 0 & 0 & 0 & 0 \\
 0 & 0 & 1 & 0 \\
 0 & 0 & -1 & 0 \\
 0 & 0 & 0 & 0
\end{array} \right)
\qquad
D_L=\left( \begin{array}{cc}
0 & \beta\\
0 & -\beta
\end{array} \right)\;.
\end{equation}

If we denote by $Q_t$ the number of particle having entered the system at time $t$ since $t=0$ and by $N_t$ the number of particle present in the system, one can define the stationary density of particles $\rho$ and current $J$ as 
\begin{eqnarray}
J=\lim_{t \rightarrow \infty} \frac{\langle Q_t \rangle}{t}\\
\rho=\lim_{t \rightarrow \infty} \frac{\langle N_t \rangle}{L} \;,
\end{eqnarray}
for which the values in the three phases are given in figure~\ref{phase_diagram}.

In order to study the large deviations of the current, one introduces the deformed Markov matrix $M(\gamma)$, obtained by multiplying by $\rme^\gamma$ in $M$ all the coefficient corresponding to a transition in which a particle enters the system. One can then show that the moment generating function of $Q_t$ is given by
\begin{equation}
\langle \rme^{\gamma Q_t} \rangle = \sum_{\mathcal{C}}\langle\mathcal{C}|\rme^{tM(\gamma)}|P_{0}\rangle\;,
\end{equation}
with $|P_{0}\rangle=\sum_{\mathcal{C}}P_{0}(\mathcal{C})|\mathcal{C}\rangle$ gathering the probabilities of all the configurations at time $t=0$. Expanding over the eigenstates of $M(\gamma)$ then gives
\begin{equation}
\langle \rme^{\gamma Q_t} \rangle = \sum_{n}\theta_n \, \rme^{t E_n(\gamma)}\;,
\end{equation}
where the $E_n(\gamma)$ are the eigenvalues of $M(\gamma)$ and the coefficient $\theta_n$ are proportional to the overlaps of the corresponding eigenstates with the initial state $|P_{0}\rangle$ of the system. All the information on the statistics of the current at finite time is then obtained by diagonalizing the matrix $M(\gamma)$. More specifically, under the KPZ scaling $t \sim L^{3/2}$, the eigenstates contributing to the fluctuations of the current in the large system size limit are the ones with eigenvalues scaling like $E_n(\gamma) \sim L^{-3/2}$, which we compute analytically in what follows.
\end{subsection}

\begin{subsection}{Cumulant generating function of the current in the large system size limit}
Using the matrix ansatz approach, Lazarescu and Mallick showed in \cite{LM2011.1} that the cumulant generating function $E(\gamma)$ of $Q_{t}$, defined by
\begin{equation}
E(\gamma) = \lim_{t \rightarrow \infty} \frac{1}{t} \log \langle \rme^{\gamma Q_t} \rangle \label{def_E}
\end{equation}
has the following parametric expression:
\begin{eqnarray}
\gamma & =-\sum _{k=1}^{\infty } \frac{C_k }{k}Z^k \label{mu_ck}\;,\\
E & =  -\sum _{k=1}^{\infty } \frac{D_k }{k}Z^k \label{E_dl}\;.
\end{eqnarray}
The coefficients of the series above are given by the contour integrals
\begin{eqnarray}
 C_k & = \frac{1}{2}\oint _{\{0,a,b\}}\frac{\dd{z} }{2 i \pi} \frac{F(z)^k}{z}\\
 D_k & = \frac{1}{2}\oint _{\{0,a,b\}}\frac{\dd{z} }{2 i \pi} \frac{F(z)^k}{(1+z)^2}\;,
\end{eqnarray}
where
\begin{equation}
 F(z) =-\frac{\left(1-z^2\right)^2 (z+1)^{2 L}}{(1-a z) (z-a) (1-b z) (z-b) z^L}
\end{equation}
and
\begin{equation}
a= \frac{1}{\alpha} - 1
\;,\qquad
b=\frac{1}{\beta} -1\;. \label{def_ab}
\end{equation}
Equivalently, one has after the resummation of the series (\ref{mu_ck}), (\ref{E_dl})
\begin{eqnarray}
  \gamma & = \frac{1}{2}\oint _{\{0,a,b\}}\frac{\dd{z} }{2 i \pi  z}\log (1-Z F(z))\;,\\
 E & = \frac{1}{2}\oint _{\{0,a,b\}}\frac{\dd{z}}{2 i \pi  (1+z)^2} \log (1-Z F(z))\;.
\end{eqnarray}

Scaling the rates $\alpha$ and $\beta$ as in (\ref{scaling_AB}) and taking for the integration contours the unit circle $|z|=1$, we observe that the contour integrals are dominated at large $L$ by $z$ close to $1$, with $z-1\sim L^{-1/2}$. Making the change of variable $z=1+\rmi y/\sqrt{L}$ ($\sqrt{A},\sqrt{B}>0$ are necessary at this step, otherwise the poles $a$ and $b$ end up on the wrong side of the contour), we obtain the asymptotics
\begin{eqnarray}
\fl
\gamma \simeq  -\frac{1}{4 \pi  \sqrt{L}}\int _{-\infty }^{\infty } \dd{y} \left(1+ \frac{y^2}{3 L}\right) \frac{ \left(\frac{16 i \sqrt{A}}{y-4 i \sqrt{A}}-\frac{16 i \sqrt{A}}{y+4 i \sqrt{A}}+\frac{16 i \sqrt{B}}{y-4 i \sqrt{B}}-\frac{16 i \sqrt{B}}{y+4 i \sqrt{B}}+2 y^2+8\right)}{\frac{e^{\frac{y^2}{4}} ((4 \sqrt{A}-i y) (4 \sqrt{A}+i y) (4 \sqrt{B}-i y) (4 \sqrt{B}+i y))}{ Z y^2}-4} \\
\fl
E \simeq -\frac{1}{4 \pi  \sqrt{L}}\int _{-\infty }^{\infty } \dd{y} \left(\frac{1}{4}+\frac{y^2}{16 L}\right) \frac{ \left(\frac{16 i \sqrt{A}}{y-4 i \sqrt{A}}-\frac{16 i \sqrt{A}}{y+4 i \sqrt{A}}+\frac{16 i \sqrt{B}}{y-4 i \sqrt{B}}-\frac{16 i \sqrt{B}}{y+4 i \sqrt{B}}+2 y^2+8\right)}{\frac{e^{\frac{y^2}{4}} ((4 \sqrt{A}-i y) (4 \sqrt{A}+i y) (4 \sqrt{B}-i y) (4 \sqrt{B}+i y))}{Z y^2}-4}\;.
\end{eqnarray}

Finally, we take the limit $B\to\infty$ in order to probe the transition line between the low density and the maximal current phase. Setting \footnote{the factor $B^2(1+A^2)$ ensures that the expressions are finite in both the $A \rightarrow 0$ and the $A \rightarrow \infty$ limits. The additional coefficient $v_0^A$ in the exponential ensures, for convenience, that the branch points of the functions $\chi^A$ and $\eta^A$ obtained \emph{infra} are on the imaginary axis of the complex plane.}
\begin{equation}
 Z = -16B^2 (1+A^2)\rme^{v+v_0^A}
\end{equation}
where
\begin{equation}
\label{def_v0}
v_0^A = \frac{2\sqrt{A}}{\sqrt{A}+\sqrt{A+4}}-2\log\Big(\frac{2\sqrt{A+4}}{\sqrt{A}+\sqrt{A+4}}\Big)\;,
\end{equation}
we define the functions
\begin{eqnarray}
  \eta^A(v)  & = \int_{-\infty}^{\infty} \dd{y} \frac{(A+4) y^2 \left(4 A \left(y^2-4\right)+y^4\right)}{2 \left(4 A+y^2\right) \left(\left(4 A+y^2\right) e^{-v-v_0^A+\frac{y^2}{4}}+(A+4) y^2\right)}\;, \label{def_eta}\\
  \chi^A(v) & =   \frac{1}{16} \int_{-\infty}^{\infty} \dd{y} \frac{(A+4) y^4 \left(4 A \left(y^2-4\right)+y^4\right)}{2 \left(4 A+y^2\right) \left(\left(4 A+y^2\right) e^{-v-v_0^A+\frac{y^2}{4}}+(A+4) y^2\right)}\;. \label{def_chi}
\end{eqnarray}
The generating function of the cumulants of the current in the limit $L \rightarrow \infty$ then has the asymptotics
\begin{eqnarray}
 & \gamma \simeq \frac{\eta^A(v)}{4\pi \sqrt{L}}\;, \label{asymptotic_gamma}\\
 & E -\frac{\gamma}{4} \simeq \frac{\chi^A(v)}{12\pi L^{3/2}}\;. \label{asymptotic_E}
\end{eqnarray}
The parameter $A>0$ describes the full crossover between the maximal current phase $A \rightarrow \infty$ and the boundary with the low density phase $A \to 0$.
\end{subsection}

\begin{subsection}{Spectral gaps in the crossover regime}
The generating function $E(\gamma)$ defined by (\ref{def_E}) is the eigenvalue of the deformed Markov matrix $M(\gamma)$ with largest real part. We claim that all other eigenvalues contributing to the KPZ universal regime in the crossover phase (namely, those scaling like $L^{-3/2}$), for $A>0$, can be obtained by similar parametric expressions as (\ref{asymptotic_gamma}), (\ref{asymptotic_E}),
\begin{eqnarray}
 & \gamma \simeq \frac{\eta_P^A(v)}{4\pi \sqrt{L}}\;, \label{asymptotic_gamma_P}\\
 & E -\frac{\gamma}{4} \simeq \frac{\chi_P^A(v)}{12\pi L^{3/2}}\;. \label{asymptotic_E_P}
\end{eqnarray}
where the functions $\eta^A_P$ and $\chi^A_P$, defined by (\ref{full_analytic_continuation eta}) and (\ref{full_analytic_continuation chi}), are other branches of the functions $\eta^A$ and $\chi^A$ above obtained by analytic continuations. These branches are indexed by finite sets of integer $P \subset \mathbb{Z}$, which also label the corresponding eigenstates.

\begin{figure}
	\begin{center}
	\includegraphics[scale=1.1]{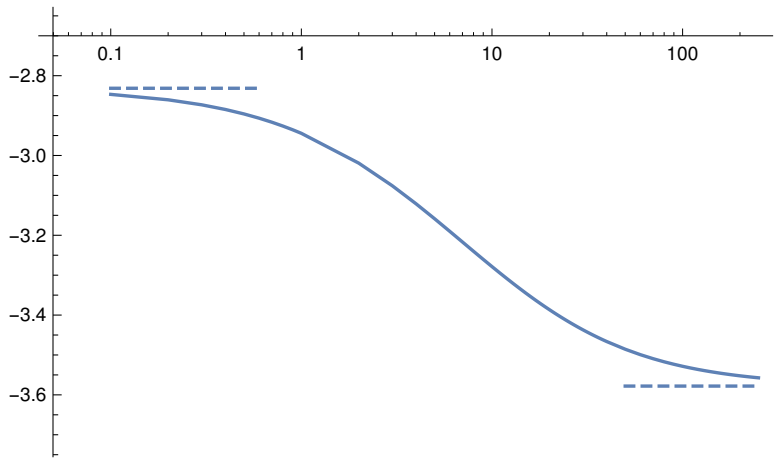}
	\caption{Plot of the spectral gap $e_1=\chi_{\{-1,1\}}^A(v)$ with $v$ solution of $\eta_{\{-1,1\}}^A(v)=0$, corresponding to a vanishing rescaled fugacity $\gamma\sqrt{L}$. Dashed lines indicate the asymptotic values of $e_1$ when $A \rightarrow 0,\infty$.}
	\label{fig gap}
	\end{center}
\end{figure}

In particular, the gap, defined as the eigenvalue with second largest real part, is obtained for the set $P=\{-1,1\}$. With the parameter $v$ fixed such that $\eta_{\{-1,1\}}^A(v)=0$ corresponding to a vanishing rescaled fugacity $\gamma\sqrt{L}$, the function $\chi_{\{-1,1\}}^A(v)$ interpolates between $-2.8315431506$ when $A=0$ and $-3.5780646645$ when $A\to\infty$, which correspond to the values obtained respectively in \cite{P2014.1} for the first gap with zero momentum \footnote{The eigenvalue with second largest real part of the periodic TASEP, computed in \cite{GS1992.1}, has non-zero momentum and is thus not recovered here.} of periodic TASEP at half-filling (up to a factor $8\sqrt{2}$) and in \cite{GP2020.1} for open TASEP in the maximal current phase, see figure~\ref{fig gap}. 
\end{subsection}

\end{section}

\begin{section}{Analytic continuations of the large deviation functions} \label{sec_analytic_continuation}
In this section we obtain the explicit expression of the analytic continuations of the functions $\eta^A$ and $\chi^A$. Some technical details about (generalized) Lambert functions are gathered in \ref{appendix_a}.
 
\begin{subsection}{Translation and analytic continuation}
\label{translation}
Let $\psi$ be the following monotonic function of $A>0$
\begin{equation}
\psi (A) = - \sqrt{A}\sqrt{A+4} -4\log\Big(\frac{\sqrt{A}+\sqrt{A+4}}{2}\Big)\;,
\end{equation}
which decreases from $\psi(0)=0$ to $\lim_{A\to\infty}\psi(A)=-\infty$.

We consider in this section the space $\mathcal{F}_{A}$ of functions analytic in the domain
\begin{equation}
\label{D}
\fl
\mathbb{D}^A=\mathbb{C}\setminus(\rmi(-\infty,-\pi]\cup\rmi[\pi,\infty) \cup (\psi (A) +\rmi(-\infty,-\pi])) \cup (\psi (A) +\rmi[\pi,\infty))  )\;,
\end{equation}
see figure \ref{figure_D}, which may be continued analytically on any path avoiding the points of $\mathcal{S}=(2\rmi\pi(\mathbb{Z}+1/2))\cup(\psi (A) +2 \rmi \pi(\mathbb{Z}+1/2))$ (i.e. such functions or any of their analytic continuations must not have branch points outside $\mathcal{S}$).

\begin{figure}
\begin{center}
\includegraphics[scale=0.5]{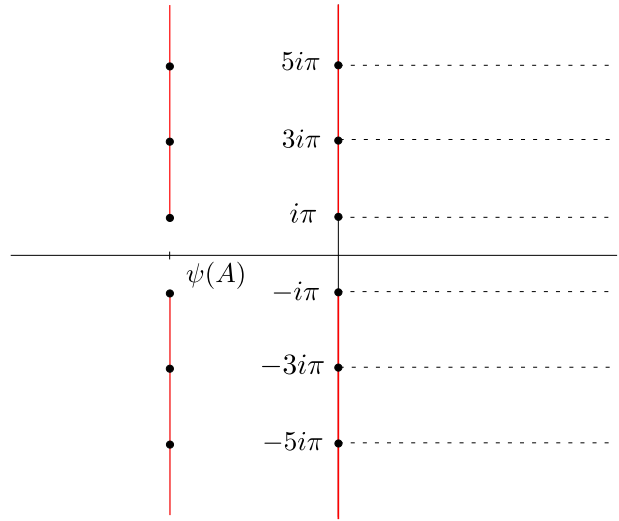}
\end{center}
\caption{Branch cuts of the functions $\eta^A$ and $\chi^A$ (dashed lines)  defined in (\ref{def_eta}) and (\ref{def_chi}), and lines excluded from the domain $\mathbb{D}^A$ (solid red lines) which are the cuts of functions $\eta^A_\emptyset$ and $\chi^A_\emptyset$ defined in (\ref{def_eta_empty}) and (\ref{def_chi_empty}).}
\label{figure_D}
\end{figure}
 
Due to the nature of the branch cuts and possible branch points required for functions in $\mathcal{F}_{A}$, we observe that $v\mapsto f(v+2\rmi\pi)$ may be extended to a function in $\mathcal{F}_{A}$ if $f\in\mathcal{F}_{A}$. More precisely, we define translation operators acting on functions $f\in\mathcal{F}_{A}$ by
\begin{eqnarray}
 \mathcal{T}^{n}_{\text{r}} f(v) = f(v+2 \rmi \pi n) & \quad  \text{if} \quad \Re v > 0\\
 \mathcal{T}^{n}_{\text{m}} f(v) = f(v+2 \rmi \pi n) & \quad \text{if} \quad \psi (A) < \Re v < 0 \\
 \mathcal{T}^{n}_{\text{l}} f(v) = f(v+2 \rmi \pi n) & \quad \text{if} \quad \Re v < \psi (A)\;,
\end{eqnarray}
where $n\in\mathbb{Z}$, and the indices stand for right ($\text{r}$), middle ($\text{m}$) and left ($\text{l}$).

Similarly, analytic continuations of any $f\in\mathcal{F}_{A}$ across any cut $2\rmi\pi(n-1/2,n+1/2)$ or $\psi(A)+2\rmi\pi(n-1/2,n+1/2)$, $n\in\mathbb{Z}^{*}$ gives another function of $\mathcal{F}_{A}$, while crossing the segments $2\rmi\pi(-1/2,1/2)$ or $\psi(A)+2\rmi\pi(-1/2,1/2)$ leave the function unchanged. For the cuts $2\rmi\pi(n-1/2,n+1/2)$, we thus define analytic continuation operators $\mathcal{A}^n_{0,\text{l}}$ and $\mathcal{A}^n_{0,\text{r}}$, $n\in\mathbb{Z}$, such that $\mathcal{A}^n_{0,\text{l}}f$ and $\mathcal{A}^n_{0,\text{r}}f$ are the functions obtained by analytic continuation along a path crossing the cut $2\rmi \pi (n-1/2, n+1/2)$, respectively from the left or from the right of the cut. For the cuts $\psi(A)+2\rmi\pi(n-1/2,n+1/2)$, we also define analytic continuation operators $\mathcal{A}^n_{\psi,\text{l}}$ and $\mathcal{A}^n_{\psi,\text{r}}$, $n\in\mathbb{Z}$, such that $\mathcal{A}^n_{\psi,\text{l}}f$ and $\mathcal{A}^n_{\psi,\text{r}}f$ are the functions obtained by analytic continuation along a path crossing the cut $\psi(A)-2\rmi \pi (n-1/2, n+1/2)$, respectively from the left or from the right of the cut.

We observe that the translation and analytic continuation operators satisfy the following relations:
\begin{eqnarray}
\mathcal{T}_{\text{r}}^{-n} \mathcal{A}^{m}_{0, \text{l}} \mathcal{T}_{\text{m}}^{n}=\mathcal{A}^{m+n}_{0, \text{l}} \nonumber\\ 
\mathcal{T}_{\text{m}}^{-n} \mathcal{A}^{m}_{0, \text{r}} \mathcal{T}_{\text{r}}^{n}=\mathcal{A}^{m+n}_{0, \text{r}}\\
\mathcal{T}_{\text{m}}^{-n} \mathcal{A}^{m}_{\psi, \text{l}} \mathcal{T}_{\text{l}}^{n}=\mathcal{A}^{m+n}_{\psi, \text{l}}\nonumber\\ 
\mathcal{T}_{\text{l}}^{-n} \mathcal{A}^{m}_{\psi, \text{r}} \mathcal{T}_{\text{m}}^{n}=\mathcal{A}^{m+n}_{\psi, \text{r}}\;.\nonumber
 \end{eqnarray}
Furthermore, the analyticity of functions in $\mathcal{F}_{A}$ on the band $-\pi < \Im v < +\pi$ imply that $\mathcal{A}^{0}_{0,\text{l}}$, $\mathcal{A}^{0}_{0,\text{r}}$, $\mathcal{A}^{0}_{\psi,\text{l}}$ and $\mathcal{A}^{0}_{\psi,\text{r}}$ are all equal to the identity operator. The relations above for $m=0$, then imply that the analytic continuation operators can be expressed in terms of translation operators as
\begin{eqnarray}
&& \mathcal{A}^{n}_{0,\text{l}}=\mathcal{T}_{\text{r}}^{-n}\mathcal{T}_{m}^{n}\nonumber\\
&& \mathcal{A}^{n}_{0,\text{r}}=\mathcal{T}_{\text{m}}^{-n}\mathcal{T}_{\text{r}}^{n} \label{algebra_AT}\\ 
&& \mathcal{A}^{n}_{\psi,\text{l}}=\mathcal{T}_{\text{m}}^{-n}\mathcal{T}_{\text{l}}^{n}\nonumber\\
&& \mathcal{A}^{n}_{\psi,\text{r}}=\mathcal{T}_{\text{l}}^{-n}\mathcal{T}_{\text{m}}^{n}\;.\nonumber 
\end{eqnarray}
\end{subsection}

\begin{subsection}{functions $\chi_\emptyset^A$ and $\eta_\emptyset^A$}
The functions $\eta^{A}$ and $\chi^{A}$ defined in (\ref{def_eta}) and (\ref{def_chi}) have branch cuts corresponding to the value of $v$ for which their respective integrands $f^{A,v}_\eta$ and $f^{A,v}_\chi$, given by
\begin{eqnarray}
f^{A,v}_\eta(y) & = \frac{(A+4) y^2 \left(4 A \left(y^2-4\right)+y^4\right)}{2 \left(4 A+y^2\right) \left(\left(4 A+y^2\right) e^{-v-v_0^A+\frac{y^2}{4}}+(A+4) y^2\right)}\\
f^{A,v}_\chi(y) & = \frac{y^2}{16} f^{A,v}_\eta(v)\;,
\end{eqnarray}
have poles $y\in\mathbb{R}$.

Apart from the points $\pm 2 \rmi \sqrt{A}\notin\mathbb{R}$, the poles of $f^{A,v}_\eta$ and $f^{A,v}_\chi$ are the solutions of
\begin{equation}
\left(4 A+y^2\right) e^{-v-v_0^A+\frac{y^2}{4}}+(A+4) y^2 = 0
\end{equation}
The solutions of this equation are given by the functions $y_j^A(v)$, defined in \ref{appendix_a} in terms of modified Lambert functions. More precisely, when $v \in \mathbb{R}^+ + 2 \rmi \pi (n + 1/2)$, the four functions $\pm y^A_n(v)$ and $\pm y^A_0(v)$ take real values. The functions $y^A_j$ defined in \ref{appendix_a}, having their branch point in $\mathcal{S}=(2\rmi\pi(\mathbb{Z}+1/2))\cup(\psi (A) +2 \rmi \pi(\mathbb{Z}+1/2))$, belong to the space $\mathcal{F}_{A}$ defined in the previous section. The branch points on the lines $\Re v = 0$ and $\Re v = \psi (A) = -\log \frac{\beta_1^A}{\beta_2^A}$ correspond to the branch points $\beta_1^A$ and $\beta_2^A$ of the functions $\Omega_j^A$ defined in \ref{appendix_a}.

We now construct determinations $\chi_\emptyset^A\in\mathcal{F}_{A}$ and $\eta_\emptyset^A\in\mathcal{F}_{A}$ of the functions $\eta^A$ and $\chi^A$ with vertical branch cuts, by computing the analytic continuation of $\eta^A$ and $\chi^A$ along the paths $v+2\rmi \pi t$, $t \in \mathbb{R}$, starting from the strip $-\pi<\Im v<\pi$. By continuously deforming the integration contour in expressions (\ref{def_eta}) and (\ref{def_chi}) to avoid the poles, one shows that the discontinuity of $\varphi^A(v) = \eta^A (v)$ or $\chi^A (v)$ when crossing the cut $\mathbb{R}_+ + 2\rmi \pi (j+1/2)$ is given by the sum of the residues of $\varphi^A (v)$ at these poles
\begin{eqnarray}
\fl
\lim_{\epsilon \rightarrow 0} \varphi^A(v + \rmi \epsilon) - \varphi^A(v-\rmi \epsilon) = 2 \rmi \pi \Big( \mathrm{res}(f^{A,v}_{\eta | \chi} (v),y^A_j(v)) - \mathrm{res}(f^{A,v}_{\eta | \chi} (v),-y^A_0(v))\\
\hspace{35mm} + \mathrm{res}(f^{A,v}_{\eta | \chi} (v),y^A_0(v)) - \mathrm{res}(f^{A,v}_{\eta | \chi} (v),-y^A_j(v)) \Big)\;, \nonumber
\end{eqnarray}
where $f^{A,v}_{\eta | \chi} (v)$ stands for either $f^{A,v}_{\eta} (v)$ or $f^{A,v}_{\chi} (v)$. Computing these residues, we find that the versions of the function $\eta^A$ and $\chi^A$ coinciding with $\eta^A_\emptyset$ and $\chi^A_\emptyset$ in the strip $-\pi<\Im v<\pi$ and belonging to the space $\mathcal{F}_{A}$ defined in the previous section, have the following expressions for $\Re v >0$:
\begin{eqnarray}
\eta^A_\emptyset (v) = \eta^A(v) - 2 i \Bigg(\frac{1-(-1)^{[\Im\frac{v}{2 \pi}]}}{2} \, y^A_0 (v) + \sum_{j \in B_{[\Im \frac{v}{2\pi}]}} y^A_j (v) \Bigg) \label{def_eta_empty}\\
\chi^A_\emptyset (v) = \chi^A(v) - 2 i \Bigg(\frac{1-(-1)^{[\Im\frac{v}{2 \pi}]}}{2} \, y^A_0 (v)^3 + \sum_{j \in B_{[\Im \frac{v}{2\pi}]}} y^A_j (v)^3 \Bigg)\;. \label{def_chi_empty}
\end{eqnarray}
The sets $B_n$ are defined in (\ref{B}), and $[\,\cdot\,]$ indicates rounding to the closest integer. For $\Re v <0$, $\eta^A_\emptyset (v)$ and $\chi^A_\emptyset (v)$ are respectively equal to $\eta^A(v)$ and $\chi^A(v)$.

\end{subsection}

\begin{subsection}{Analytic continuation of functions $\eta^A_\emptyset$ and $\chi^A_\emptyset$}\label{ac_chi_eta}
We now move on to the computation of the full analytic continuation of the functions $\eta^A_\emptyset$ and $\chi^A_\emptyset$ across their cuts $2\rmi \pi (n-1/2, n+1/2)$ and $\psi(A)+2\rmi \pi (n-1/2, n+1/2)$, $n\in\mathbb{Z}^{*}$. Since the overall structure of their analytic continuation is similar, we will denote by $\varphi_\emptyset^A$ any of the two function, and by $\lambda_j^A$ the functions $y_j^A$ or $(y_j^A)^3$ depending on the expression of the residues appearing in expressions (\ref{def_eta_empty}) and (\ref{def_chi_empty}).
 
 Since the function $\varphi^A(v)$ is analytic on the strips $\{v \in \mathbb{C}, 2 \rmi \pi (n-1/2) < \Im v < \rmi \pi (n+1/2) \}$, the definition (\ref{def_eta_empty}) and (\ref{def_chi_empty}) leads to the expression
 \begin{eqnarray}
  \mathcal{A}_{0,l}^n~\varphi_\emptyset^A (v) = \varphi_\emptyset^A (v) + 2 i \Bigg(\frac{1-(-1)^{[\Im\frac{v}{2 \pi}]}}{2} \, \lambda^A_0 (v) + \sum_{j \in B_{[\Im \frac{v}{2\pi}]}} \lambda^A_j (v) \Bigg)
 \end{eqnarray}
applying the relations obtained in \ref{appendix_wj_yj} to the functions $\lambda_j$, one can similarly compute the analytic continuation of $\varphi^A(v)$ from the other side of the cut
\begin{equation}
 \mathcal{A}_{0,r}^n~\varphi_\emptyset^A (v) = \varphi_\emptyset^A (v) + 2 i \Bigg(\frac{1+(-1)^{[\Im\frac{v}{2 \pi}]}}{2} \, \lambda^A_n (v) + \sum_{j \in B_{[\Im \frac{v}{2\pi}]}} \lambda^A_j (v) \Bigg)
\end{equation}
$\phi_\emptyset^A(v)$ being analytic for $\Re v < 0$, the operators $\mathcal{A}^n_{\psi, \mathrm{l}}$, $\mathcal{A}^n_{\psi, \mathrm{r}}$ act as identity.

We see that in any case the continuation across its cuts of $\varphi^A_\emptyset$ is obtained in the form of a function $\varphi^A_P$, where $P$ is a set of integer indices indicating which functions $\lambda_j^A$ are added up to $\varphi^A_\emptyset$. Explicitly 
\begin{eqnarray}
\label{full_analytic_continuation eta}
\eta^A_P (v) = \varphi^A_\emptyset(v) + 2 \rmi \sum_{j \in P} y^A_j(v) \;, \\
\label{full_analytic_continuation chi}
\chi^A_P (v) = \chi^A_\emptyset(v) + 2 \rmi \sum_{j \in P} ( y^A_j(v) )^3 \;.
\end{eqnarray}
The sets $P$ parametrize the several branches of the analytic continuation of $\varphi_A$. Using the expressions derived in \ref{appendix_wj_yj}, we can express the analytic continuation and translation operators on functions $\varphi_P^A$ as set-theoretic operations on sets $P$. Writing
\begin{eqnarray}
 \mathcal{T}^n_{0 | \psi, \mathrm{l}|\mathrm{r}} \varphi^A_P = \varphi^A_{T^n_{0 | \psi,\mathrm{l}|\mathrm{r}} P} \\
 \mathcal{A}^n_{0 | \psi, \mathrm{l}|\mathrm{r}} \varphi^A_P = \varphi^A_{A^n_{0 | \psi,\mathrm{l}|\mathrm{r}} P}\;,
\end{eqnarray}
where $\mathrm{l}|\mathrm{r}$ stands for either $\mathrm{l}$ or $\mathrm{r}$ and $0 | \psi$ for either $0$ or $\psi$, the action of the translation operators read
\begin{eqnarray}
\fl
 T_\text{m}^{-n} P = P + n\\
 \fl
  T_{\text{r}}^{-n}P=
\left\{
	\begin{array}{ll}
		((P+n) \setminus C_{n}) \cup (B_{n} \setminus (P+n+\sgn~n)) \cup \{0\}
		&
		\begin{array}{ll}
			0\in P \;\&\; n\;\text{even}\\
			0\notin P \;\&\; n\;\text{odd}
		\end{array}\\[5mm]
		((P+n) \setminus C_{n}) \cup (B_{n} \setminus (P+n+\sgn~n))
		&
		\begin{array}{ll}
			0\in P \;\&\; n\;\text{odd}\\
			0\notin P \;\&\; n\;\text{even}
		\end{array}
	\end{array}
\right.\\
\fl
 T_{\text{l}}^{-n}P=
\left\{
	\begin{array}{ll}
		((P+n) \setminus C_{n}) \cup ((P \cap B_{-n})+n+\sgn~n) \cup \{0\}  & \quad 0\in P \\
		((P+n) \setminus C_{n}) \cup ((P \cap B_{-n}) +n+\sgn~n) &  \quad 0\notin P

	\end{array}
\right.\;.
\end{eqnarray}
where the sets $B_n$ are defined as
\begin{equation}
\label{B}
B_n = \left\{
\begin{array}{cl}
	\{1,\ldots,n\} \quad & \text{if} \quad n > 0\\
	\emptyset \quad & \text{if} \quad n = 0\\
	\{n,\ldots,-1\} \quad & \text{if} \quad n < 0
\end{array}
\right.\;,
\end{equation}
and $C_n = B_n \cup \{0\}$. Using relations (\ref{algebra_AT}), one can finally derive the actions of analytic continuation operators on $\varphi^A_P$
\begin{eqnarray}
\fl
A^{n}_{0,\text{l}}P=
\left\{
	\begin{array}{ll}
		(P \setminus C_{n}) \cup (B_{n} \setminus (P+\sgn~n)) \cup \{0\}
		&
		\begin{array}{ll}
			n\in P \;\&\; n\;\text{even}\\
			n\notin P \;\&\; n\;\text{odd}
		\end{array}\\[5mm]
		(P \setminus C_{n}) \cup (B_{n} \setminus (P+\sgn~n))
		&
		\begin{array}{ll}
			n\in P \;\&\; n\;\text{odd}\\
			n\notin P \;\&\; n\;\text{even}
		\end{array}
	\end{array}
\right.\label{set_continuation1} \\ 
\fl
A^{n}_{0,\text{r}}P=
\left\{
	\begin{array}{ll}
		(P \setminus C_{n}) \cup (C_{n} \setminus (P-\sgn~n)) \cup \{n\}
		&
		\begin{array}{ll}
			0\in P \;\&\; n\;\text{even}\\
			0\notin P \;\&\; n\;\text{odd}
		\end{array}\\[5mm]
		(P \setminus C_{n}) \cup (C_{n-\sgn~n} \setminus (P-\sgn~n))
		&
		\begin{array}{ll}
			0\in P \;\&\; n\;\text{odd}\\
			0\notin P \;\&\; n\;\text{even}
		\end{array}
	\end{array}
\right.\!\!.\\
\fl
A^{n}_{\psi, \text{l}} P = 
\left\{
	\begin{array}{ll}
		(P\setminus C_{n}) \cup ((P \cap B_{n})-\sgn~n) \cup \{n\}  & \quad 0\in P \\
		(P\setminus C_{n}) \cup ((P \cap B_{n})-\sgn~n)  &  \quad 0\notin P

	\end{array}
\right.\;,\\
\fl
A^{n}_{\psi, \text{r}} P = 
\left\{
	\begin{array}{ll}
		(P\setminus C_{n}) \cup ((P \cap C_{n-\sgn~n})+\sgn~n) \cup \{0\}  & \quad n\in P \\
		(P\setminus C_{n}) \cup ((P \cap C_{n-\sgn~n})+\sgn~n)  &  \quad n\notin P

	\end{array}
\right.\;. \label{set_continuation2}
\end{eqnarray}
These formulas characterize how sheets of the Riemann surface $\mathcal{R}^{A}$ defined below are glued to one another.

The parametric equations (\ref{asymptotic_gamma_P}) and (\ref{asymptotic_E_P}) being invariant under the change of variable $v \rightarrow v + 2 k \rmi \pi$, each eigenstate is in fact characterized by an orbit of sets $P$ under the action of operators $T_\mathrm{r}^n$. One can show (see \ref{appendix_riemann}) that in the orbit of any finite set $P \subset \mathbb{Z}$ there is a single set $P^\star$ that has the same number of positive and negative element. Moreover the collection of all $P \subset \mathbb{Z}$ such that $0 \notin P^\star$ is globally stable under the action of the analytic continuation operators defined above. We conjecture that a complete set of eigenstates contributing at large $L$ to the KPZ regime with finite boundary parameter $A$ is given by all such sets $P^{*}\subset\mathbb{Z}^{*}$ having the same number of negative and positive elements.

\end{subsection}

\begin{subsection}{Limit cases $A \rightarrow \infty$ and $A \rightarrow 0$}
We now consider the limit cases where $A$ is taken to infinity and $0$, which are expected to reduce to the expressions already known for the first excitations of the periodic TASEP at half-filling when $A\to0$ and of the open TASEP in the maximal current phase when $A \rightarrow \infty$. 

Let us first consider the $A\to0$ case. Making the change of variable $y=2\sqrt{t}$ in the integrals (\ref{def_eta}) and (\ref{def_chi}) and using $\int_{0}^{\infty}\rmd t\,\frac{t^{s-1}}{\rme^{t-v}-1}=\Gamma(s)\Li_{s}(\rme^{v})$ with $\Gamma$ the Euler gamma function and $\Li_s$ the polylogarithm defined for $s \in \mathbb{C}$, $|z| < 1$ as
\begin{equation}
\Li_s(z) = \sum_{k=1}^\infty \frac{z^k}{k^s}\;,
\end{equation}
we obtain in terms of the variable $c=v/(2\pi)$
\begin{eqnarray}
\eta^0(c) & = 2\sqrt{\pi} \, \Li_{3/2} (-\rme^{2\pi c})\\
\chi^0(c) & = 12 \sqrt{\pi} \, \Li_{5/2} (-\rme^{2 \pi c})\;.
\end{eqnarray}
The functions $\Omega^A_j$ defined in \ref{appendix_a} become $\Omega^0_j(z) = \log (-z/4) + 2\rmi \pi (j+1/2)$ in this limit, so that for $j \neq 0$, the functions $y^A_j(v)$ read
\begin{equation}
 y^0_j (c)= -\sgn (j) \sqrt{2\pi}\sqrt{c + \rmi \left(j - \sgn (j) \frac{1}{2} \right)}\;.
\end{equation}
while $y_0^A (v) \rightarrow 0$ when $A \rightarrow 0$. Thus, we recover the general expressions of first excited states eigenvalues with zero momentum obtained in \cite{P2014.1} for TASEP with periodic boundary conditions. In the notations of \cite{P2020.1}, such excited states with zero momentum correspond to identical sets of particle and hole excitations.

Similarly, taking the limit $A \rightarrow \infty$ in expressions (\ref{asymptotic_gamma})-(\ref{asymptotic_E}), one obtains the following asymptotic expressions for $E(\gamma)$
\begin{eqnarray}
 \gamma = \frac{1}{3 \pi \sqrt{L}} \int_{-\infty}^{+ \infty} \dd{y} \frac{\left(1-y^2\right) \left(3-y^2\right) \rme^{-v+y^2-1}}{y^2 \left(y^{-2} \rme^{-v+y^2-1}+1\right)^2}\\
 E - \left(\frac{1}{4L} + \frac{1}{4} \right) \gamma = \frac{1}{12\pi L^{3/2}} \int_{-\infty}^{+ \infty} \dd{y} \frac{\left(1-y^2\right) \left(3-y^2\right)}{y^{-2} \rme^{-v} \rme^{y^2-1}+1}
\end{eqnarray}
while the functions $\Omega_j^A(z)$ become equal to the Lambert $W_j$ functions as explained in \ref{appendix_a}, so that one gets finally the expression already derived in \cite{GP2020.1} for open TASEP in the maximal current phase.
\end{subsection}

\begin{subsection}{Riemann surface $\mathcal{R}^A$}
The natural domain of definition of a multivalued meromorphic function is a Riemann surface, defined by gluing together the domains of definition of the various branches of the multivalued function along branch cuts according to analytic continuation. While the partition of the Riemann surface into sheets is arbitrary and depends on the choice of branch cuts for the multivalued function, the Riemann surface itself is uniquely defined by the function on its original domain.

In the case of the open TASEP in the crossover regime studied in this paper, the functions $\eta^A_P$ and $\chi^A_P$ can be extended to the same non-compact Riemann surface (of infinite genus) $\mathcal{R}^A$, constructed by gluing together along the lines $2\rmi\pi(n-1/2,n+1/2)$ and $\psi (A) + 2\rmi\pi(n-1/2,n+1/2)$, $n\in\mathbb{Z}^{*}$ copies $\mathbb{D}^A_P$ of the domain $\mathbb{D}^A$ corresponding to the function $\eta^A_P$ or $\chi^A_P$. If we denote by $[v,P]$ the point of the sheet $\mathbb{D}^A_P$ that project itself on $v \in \mathbb{C}$, the functions $\eta^A$ and $\chi^A$ are extended into meromorphic functions $\mathfrak{X}^A$ and $\mathfrak{H}^A$ on $\mathcal{R}^A$ defined by $\mathfrak{H}^A([v,P]) = \eta^A_P (v)$ and  $\mathfrak{X}^A([v,P]) = \chi^A_P (v)$.

The connectivity of the sheets $\mathbb{D}^A_P$ is fixed by the relations (\ref{set_continuation1})-(\ref{set_continuation2}) relating the branches to one another by analytic continuation: when crossing from sheet $\mathbb{D}^A_P$ the cut $2\rmi\pi(n-1/2,n+1/2)$ or $\psi (A) + 2\rmi\pi(n-1/2,n+1/2)$ from the left or from the right side, one will end up on the sheet $\mathbb{D}^A_{A^n_{\psi | 0, r|l}P}$ with the analytic continuation operator corresponding to the chosen path.

As explained in section~\ref{ac_chi_eta}, each eigenstate is characterized by a sheet corresponding to a set $P \subset \mathbb{Z}^\star$ (called $P^{*}$ in section~\ref{ac_chi_eta}) that has the same number of positive and negative elements, $|P_-|=|P_+|$. For $0 \leq A < \infty$ all such sheets form a connected component $\mathcal{R}^A_*$ of $\mathcal{R}^A$ obtained by analytic continuations from the sheet $\mathbb{D}^A_\emptyset$. The connected component $\mathcal{R}_{0}$ containing all the sheets $\mathbb{D}^A_P$ indexed by finite sets $P$ with the same number of positive and negative elements such that $P$ contains $0$ does not appear to correspond to eigenstates of TASEP.

When $A \rightarrow \infty$, the branch cut at $\Re v = \psi (A)$ goes to infinity, and one recovers from $\mathcal{R}^A_*$ the Riemann surface constructed in \cite{GP2020.1} for the maximal current phase. This Riemann surface has in fact infinitely many connected components $\mathcal{C}_k$, $k \in \mathbb{N}$, corresponding to sets $P$ with the same excess number of elements of the same parity, $\bigg| | P_\text{odd} | - | P_\text{even} | \bigg|=2k$, see \ref{appendix_riemann}.

Considering now the limit $A \rightarrow 0$, since $\psi(A)\to0$, the two lines of branch cuts for the functions $\eta^A_P$ and $\chi^A_P$ collapse into one. Analytic continuations across this single line of branch cuts is realized by the products of operators $A^n_{0,\mathrm{r}} A^n_{\psi,\mathrm{r}}$ and $A^n_{\psi,\mathrm{l}} A^n_{0,\mathrm{l}}$. One has for any finite set $P \subset \mathbb{Z}$
\begin{equation}
 A^n_{0,\mathrm{r}} A^n_{\psi,\mathrm{r}} P =  A^n_{\psi,\mathrm{l}} A^n_{0,\mathrm{l}} P = \left\{
	\begin{array}{ll}
	P \ominus C_n &  \quad n \; \text{odd}\\
	P \ominus B_n  & \quad \text{otherwise}
	\end{array}
\right.\;,
\end{equation}
where $\ominus$ denotes the symmetric difference operation on sets $P\ominus Q=(P\cup Q)\setminus(P\cap Q)$. As shown in the previous section, $y_0^A (v) \rightarrow 0$ when $A \rightarrow 0$, hence for any $P\subset\mathbb{Z}^{*}$ one has $\eta^A_P=\eta^A_{P\cup\{0\}}$ and $\chi^A_P=\chi^A_{P\cup\{0\}}$ when $A\to0$, and both connected components $\mathcal{R}^A_0$ and $\mathcal{R}^A_*$ are identical to the Riemann surface $\mathcal{R}$ defined in \cite{P2020.1} and describing the KPZ fixed point with periodic boundary conditions.

The family of Riemann surfaces $\mathcal{R}^A$ obtained in this paper thus interpolates between the Riemann surface for the periodic KPZ fixed point from \cite{P2020.1} and the Riemann surface for the KPZ fixed-point with boundary conditions $\partial_x h(x=0,t) = +\infty$ and $\partial_x h(x=1,t) = -\infty$ from \cite{GP2020.1}.
\end{subsection}

\end{section}

\begin{section}{Numerical checks}\label{sec_numerics}
In this section we check the results obtained by analytic continuation for the eigenvalues of the deformed Markov matrix $M(\gamma)$ against numerical solutions of the Bethe ansatz equations.

We use the Bethe ansatz equations conjectured by Cramp\'{e} and Nepomechie in \cite{CN2018.1} from Baxter's $TQ$ equation. In this formalism, any eigenvalue of the matrix $M(\gamma)$ can be expressed in terms of the $L+2$ Bethe roots satisfying the following Bethe equations for $1 \leq j \leq L+2$
 \begin{equation}
\fl u_j^L(u_j+a)(u_j+b)(au_j+1)(bu_j+1)=(-1)^{L+1} \rme^{2\gamma}(1-u_j)^{2L+2}(u_j+1)^2 \prod_{k=0}^{L+1} u_k \;. \label{BAE}
 \end{equation}
where $a$ is defined defined by (\ref{def_ab}) and $b=1$ (which amounts to take the limit $B \rightarrow \infty$). The eigenvalue $E(\gamma)$ corresponding to a given set of solutions $\{u_0,...u_{L+1}\}$ is expressed as $E(\gamma) = - \frac{1}{2} \Lambda ' (1)$ where
\begin{equation}
\label{Lambda_vp}
\Lambda (x)= \frac{x^{L}(x+a)(x+b) }{\rme^{\gamma}\,Q(x)} + \rme^\gamma \, \frac{Q(0)}{Q(x)} \frac{(1-x)^{2L+2} (x+1)^2}{(ax+1)(bx+1)} \;.
\end{equation}
with $Q(x) = \prod_{j=0}^{L+1}(x - u_j)$. Setting $C = \rme^{2\gamma} \prod_{k=0}^{L+1} u_k$ and taking the $(2L+2)$-th root of (\ref{BAE}), the Bethe ansatz equations can be expressed in separated form
\begin{equation}
\label{power b e}
\frac{(1-u_{j})(1+u_{j})^{\frac{1}{L+1}}C^{\frac{1}{2L+2}}}{u^{\frac{L}{2L+2}}(u+a)^{\frac{1}{2L+2}}(u+b)^{\frac{1}{2L+2}}(1+au)^{\frac{1}{2L+2}}(1+bu)^{\frac{1}{2L+2}}}=\rme^{-\frac{2\rmi\pi k_{j}}{2L+2}}\;,
\end{equation}
where $\{k_0,...,k_{L+1}\}$ is a set of integers if $L$ is even, half-integers if $L$ is odd, defined modulo $2L+2$, and which characterizes the eigenstate. The $k_j$ that represents the momentum of the quasi-particle associated to the Bethe root $u_j$, have to be fixed prior to solving numerically the Bethe ansatz equations. The equations (\ref{power b e}) are then solved iteratively with Newton's method. In practice the values obtained for small system sizes are extrapolated using Richardson algorithm (see figure \ref{table_extrapolation}) and checked against the values obtained by analytic continuation of the cumulant generating function. 

\begin{table}
\begin{center}
\label{table_extrapolation}
$\begin{array}{|r|l|l|}
\hline
L & \quad e_1(L) & \quad e_1~~\text{extrapolated}\\\hline
10 & -2.98668 & \\
11 & -2.98662 & -2\\
12 & -2.98605 & -2.9\\
13 & -2.98520 & -2.9\\
14 & -2.98418 & -2.944\\
15 & -2.98308 & -2.9445\\
16 & -2.98195 & -2.9445\\
17 & -2.98082 & -2.9444519\\
18 & -2.97972 & -2.94445191\\
19 & -2.97864 & -2.944451909\\
20 & -2.97010 & -2.94445190868\\
&\quad\cdots&\qquad\cdots\\
25 & -2.97387 & -2.94445190868527914\\
&\quad\cdots&\qquad\cdots\\
30 & -2.97010 & -2.94445190868527913701680\\
&\quad\cdots&\qquad\cdots\\
35 & -2.96712 & -2.944451908685279137016796857\\
&\quad\cdots&\qquad\cdots\\
40 & -2.96432 & -2.9444519086852791370167968573037740\\
\hline
\end{array}$
\end{center}
\caption{Numerical evaluations of the spectral gap $e_{1}(L)=L^{3/2}E_{1}$ for $A=1$ and a fugacity $\gamma=0$. The middle column corresponds to a direct numerical solution of the Bethe equations for a system of $L$ sites. The last column is the extrapolated value for all system sizes between $L=10$ and the value of $L$ for the current line using the BST algorithm with exponent $\theta=1$ (for a presentation of numerical extrapolation in the context of TASEP, see for instance \cite{GP2020.1}). Numerical solutions of the Bethe equations are performed with 100 significant digits. Extrapolated values are truncated at the estimated order of magnitude of the error given by the BST algorithm.}
\label{gap_extrapolation}
\end{table}
The natural choice $k^{(0)}_j = j -1 - L/2$ leads to the ground state. Every excitation can then be parametrized by indicating which quasi-momenta are removed from the set of the $k_j^{(0)}$ (\emph{holes}) and which are added outside the interval $[-\frac{L}{2}, \frac{L}{2}]$ (\emph{particles}) when computing a given eigenstate. In the case of the open TASEP, the only excitations that correspond to actual eigenstates are the one having zero total momentum (which is not necessarily true in the case of the periodic TASEP treated in \cite{P2014.1}). It implies that eigenstates are defined in terms of a finite set $P \subset \mathbb{Z}^*$ having the same number of positive and negative elements, $|P_+|=|P_-|$. Each eigenstate corresponds to a choice of quasi momenta
\begin{eqnarray}
\fl
 \{k_j, 1 \leq j \leq L+2\} = & \left[ \Big\{ -\frac{L}{2}, -\frac{L}{2}+1,...,\frac{L}{2}-1,\frac{1}{2} \Big\} \setminus \Big( (\frac{L}{2} + P_-) \cup (-\frac{L}{2} + P_+) \Big)  \right]\nonumber\\
 & \cup \Big((\frac{L}{2} + P_+) \cup (-\frac{L}{2} + P_-)\Big)
\end{eqnarray}
The eigenvalues obtained are smooth functions of $A$. The first excitations are plotted in figure \ref{graph_eigenvalues}. They match perfectly with the values obtained by analytic continuation, where the set $P$ indexing the eigenstate is the same as the set indexing the branches of the functions $\chi^A$ and $\eta^A$.

\begin{figure}
\begin{center}
\includegraphics[width=150mm]{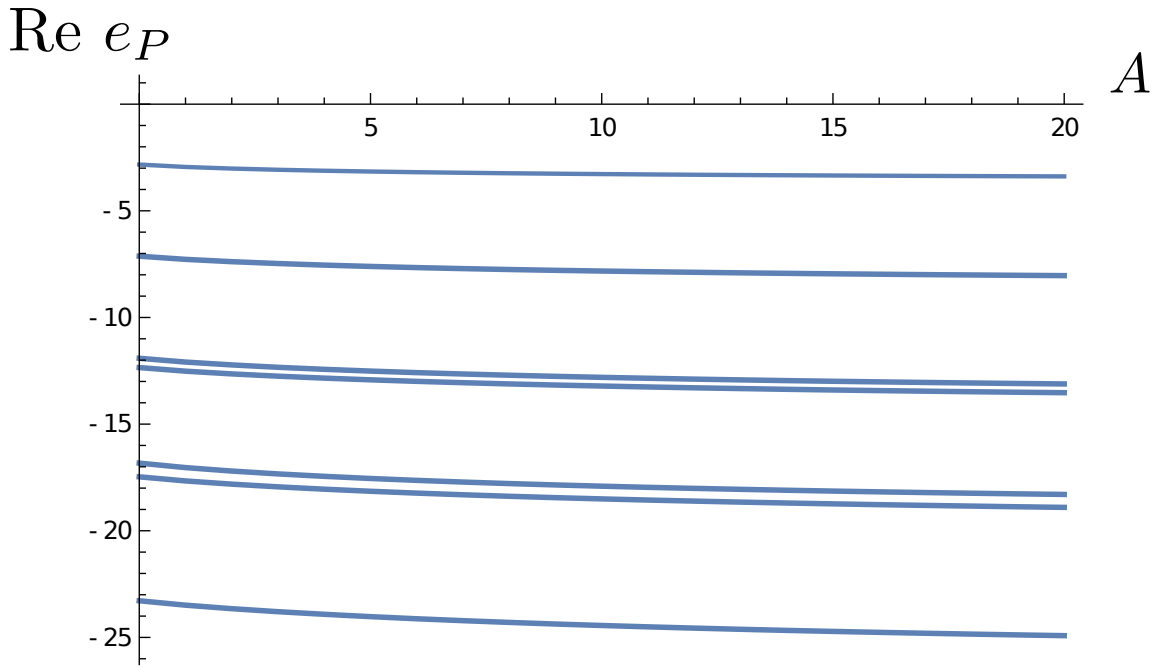}
\caption{Plot of the real part of the first eigenvalues $e_n=L^{3/2}E_n$ of $M(\gamma)$  as a function of $A$. The sets $P$ indexing these are eigenstate are,from top to bottom: $\{-1,1\}$ (gap), $\{-1,2\}$/$\{-2,1\}$, $\{-2,2\}$, $\{-1,3\}$/$\{-3,1\}$, $\{-2,-1,1,2\}$, $\{-3,2\}$/$\{-2,3\}$, $\{-3,3\}$. The alternatives $\{\ldots\}$/$\{\ldots\}$ correspond to degeneracies for $\Re e_n$.} 
\label{graph_eigenvalues}
\end{center}

\end{figure}
\end{section}

\begin{section}{Conclusion}
We have obtained in this paper an exact expression for the spectral gaps of the totally asymmetric exclusion process with open boundary conditions, with a parameter $A>0$ probing the edge of the maximal current phase. These spectral gaps are expected to correspond to relaxation times for the statistics of the KPZ height function on an interval, with an infinite slope on one side and an arbitrary finite slope on the other side. These spectral gaps are obtained from the cumulant generating function for the stationary current by analytic continuations with respect to a parameter $v$ conjugate to the fugacity of the current. Our expressions were checked with high-precision against numerical solutions of the Bethe ansatz equations.
 
The spectral gaps are associated to a family of infinite genus Riemann surfaces $\mathcal{R}^A$ by analytic continuations of the eigenstates. When the parameter $A$ is varied from $0$ to $\infty$, we observe that the Riemann surface $\mathcal{R}^A$ undergoes a continuous deformation, connecting the analytic structure of the KPZ fixed point with periodic boundary conditions and the one for the case with open boundaries with infinite slopes for the height function at both ends of the system.

Exact probability distributions for the height function at the KPZ fixed-point on a bounded interval are still missing, unlike in the periodic case where some results are known \cite{P2016.1,BL2018.1}. Our exact expressions for the spectral gaps seem to indicate that the Riemann surface approach from \cite{P2020.1,P2020.2} should also work for the probabilities in the case with open boundaries.
\end{section}

\appendix
\begin{section}{Generalized Lambert Functions} \label{appendix_a}
 
 \begin{subsection}{The Lambert $W_j$ functions}
The Lambert $W_j$ functions \cite{CGHJK1996.1} are defined as the solutions to $W_j(z) \rme^{W_j(z)} = z$ such that
\begin{equation}
W_j(z) + \log W_j(z) = \log z + 2 \rmi \pi j \;.
\end{equation}
where the logarithm function is understood as its principal determination with branch cut on $\mathbb{R}_-$. They play an important role in the computation of the cumulant generating function of the current in the maximal current phase \cite{GP2020.1}. Before introducing their analogue needed for the edge of the maximal current phase, we recall some facts about the analytic continuation of the $W_j$. 
 
The branches $W_j$, $j\in\mathbb{Z}^{*}$ are analytic in $\mathbb{C} \setminus (-\infty, 0]$ with disjoint images. For $j=0$, the function $W_0$ is analytic on $\mathbb{C} \setminus (-\infty, -\frac{1}{\rme}]$. For $j\notin\{-1,0,1\}$, the $W_j$ have a single branch point at $0$ and the branch cut is conventionally chosen to be the negative real axis $\mathbb{R}^-$. The function $W_0	$ has a single branch point at $z=-\frac{1}{e}$ and its branch cut is the half-line $(-\infty, -\frac{1}{\rme}]$. The function $W_{-1}$ has branch cut $\mathbb{R}^-$, with a branch point at $0$ and another branch point at $z=-\frac{1}{e}$ on the top side of its cut. The same goes for $W_1$, with the branch point at $z=-\frac{1}{e}$ lying on the bottom side of the cut. The analytic continuation of any $W_j$ through a cut gives another branch, as summarized in figure~\ref{analytic_continuation_lambert}, see also figure~\ref{ImageW}.
 
\begin{figure}
\includegraphics[scale=1]{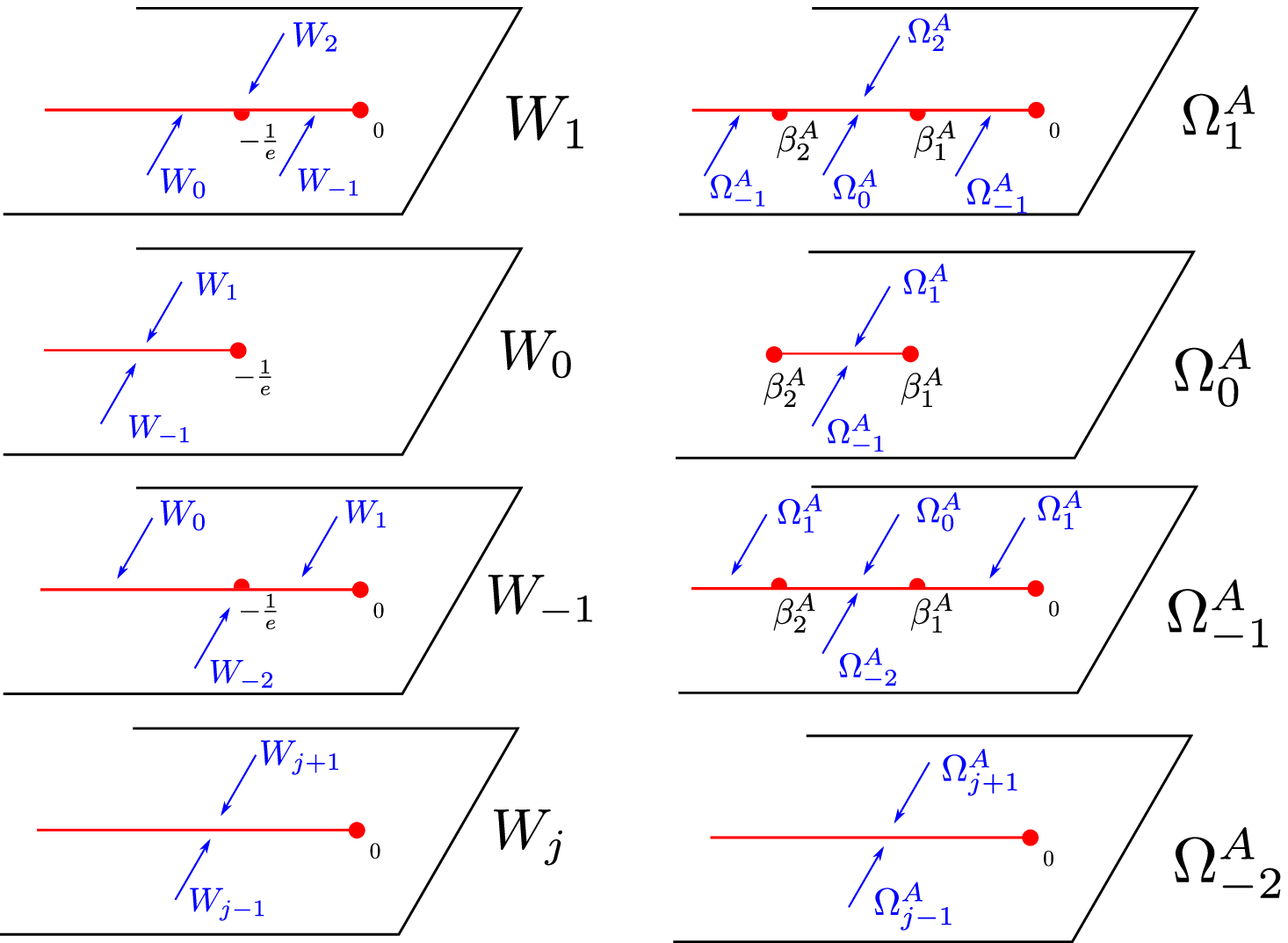}
\caption{Sketch of the analytic continuation of the Lambert $W_j$ functions (left column) and their generalized counterpart $\Omega_j^A$ (right column) in the complex plane. Blue labels indicate the function obtained by analytic continuation through the cut in the direction indicated by the corresponding arrow.}
\label{analytic_continuation_lambert}
\end{figure}

\begin{figure}
	\begin{tabular}{ll}
		\hspace{-5mm}
		\begin{tabular}{l}\includegraphics[width=75mm]{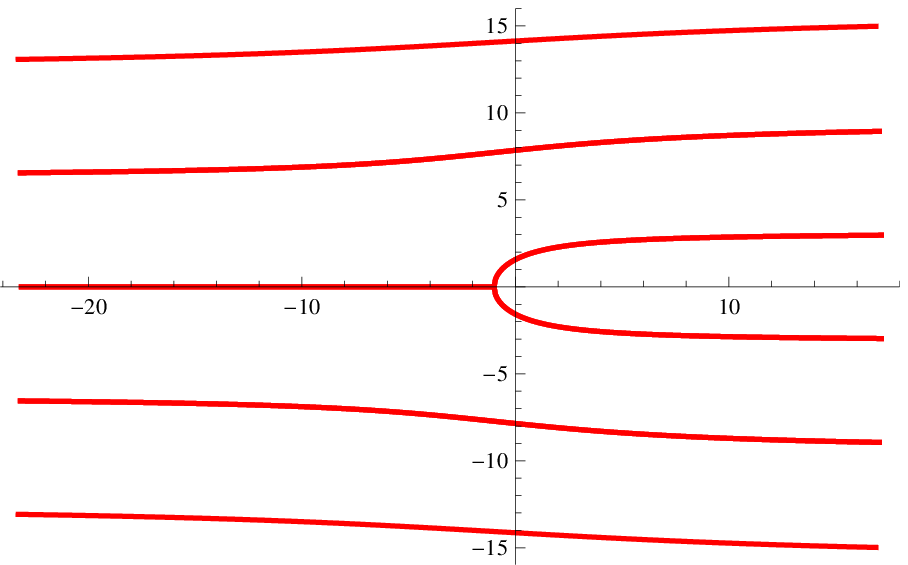}\end{tabular}
		\begin{picture}(0,0)
		\put(-30,-17){$W_{-2}$}
		\put(-30,-8){$W_{-1}$}
		\put(-30,0){$W_{0}$}
		\put(-30,8){$W_{1}$}
		\put(-30,17){$W_{2}$}
		\end{picture}
		&
		\begin{tabular}{l}\includegraphics[width=75mm]{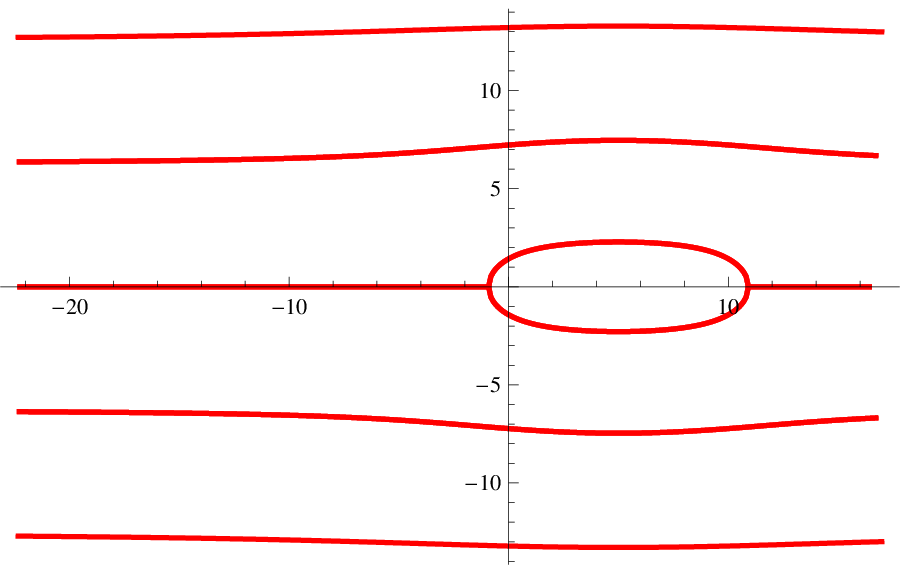}\end{tabular}
		\begin{picture}(0,0)
		\put(-30,-17.4){$\Omega_{-2}^{A}$}
		\put(-30,-8.4){$\Omega_{-1}^{A}$}
		\put(-30,0.4){$\Omega_{0}^{A}$}
		\put(-30,8.4){$\Omega_{1}^{A}$}
		\put(-30,17.4){$\Omega_{2}^{A}$}
		\put(-55,-7){\scriptsize$\Omega_{0}^{A}(\beta_{1}^{A})$}
		\put(-15,-7){\scriptsize$\Omega_{0}^{A}(\beta_{2}^{A})$}
		\put(-44,-4.5){\thinlines\vector(1,1){5}}
		\put(-13.5,-3.7){\thinlines\vector(-1,2){2}}
		\end{picture}
	\end{tabular}
	\caption{Image of the complex plane by the functions $W_{j}$ (left) and $\Omega_{j}^{A}$ with $A=10$ (right). The (red) curves correspond to the images by the $W_{j}$ and $\Omega_{j}^{A}$ of the branch cuts from figure~\ref{analytic_continuation_lambert}.}
	\label{ImageW}
\end{figure}
\end{subsection}

\begin{subsection}{Generalized Lambert functions $W_j^A$ and $\Omega_j^A$}
\label{section OmegajA}
We now introduce the generalized Lambert functions $W_j^A$ as the solutions to the equation
\begin{equation}
\rme^{W_j^A(z)} \frac{W_j^A(z)}{A-W_j^A(z)} =z \;.
\end{equation}
More precisely, for any $j \in \mathbb{Z}$, we define $W_j^A(z)$ as the only solution to
\begin{equation}
W_j^A(z) + \log \Big(\frac{W_j^A(z)}{A-W_j^A(z)} \Big) = \log z + 2\rmi \pi j
\end{equation}
with the same convention as before for the branch cut of the logarithm. It follows from the definition that for any $j \in \mathbb{Z}$
\begin{equation}
 \lim_{A \rightarrow \infty} W_j^A (z/A) = W_j(z)
\end{equation}
Given a meromorphic function $f(y)$, the branch points $z_{*}$ of the inverse function $f^{-1}$ are equal to $z_{*}=(y_{*})$ where $y_{*}$ is a solution of $f'(y_{*})=0$. For the generalized Lambert functions $W_j^A$, one has $f(y)=\rme^{y} \frac{y}{A-y}$, and we find the solutions $y_{*}=-\infty$, $y_{*}=\frac{1}{2}(A\pm\sqrt{A}\sqrt{A+4})$, corresponding to the three branch points $z_{*}=0$, $z_{*}=b_1^A$ and $z_{*}=b_2^A$, with
\begin{eqnarray}
b_1^A = & -\tfrac{1}{4}(\sqrt{A+4}-\sqrt{A})^{2}\,e^{\frac{1}{2}(A-\sqrt{A} \sqrt{A+4})}\\
b_2^A = & -\tfrac{1}{4}(\sqrt{A+4}+\sqrt{A})^{2}\,e^{\frac{1}{2}(A+\sqrt{A} \sqrt{A+4})}\;.
\end{eqnarray}
The branch points are ordered as $b_2^A<b_1^A<0$ for any $A>0$.

In the following, we set $\Omega_j^A(z) = W_j^A(\frac{z}{4+A})$ and work with the modified Lambert functions $\Omega_j^A(z)$ instead for convenience. The corresponding branch points are $0$, $\beta_{1}^A = (4+A) b_{1}^A$ and $\beta_{2}^A = (4+A) b_{2}^A$. For $j\notin\{-1,0,1\}$, we observe that $\Omega_j^A$ has a single branch point at $z=0$, and we choose the corresponding branch cut as the negative real axis. $\Omega_0^A$ only has the branch points $\beta_1^A$ and $\beta_2^A$ and we place a branch cut on the segment $[\beta_1^A, \beta_2^A]$. Finally, $\Omega_1^A$ (respectively $\Omega_{-1}^A$) has for branch points $0$, $\beta_1^A-\rmi0^{+}$ and $\beta_2^A-\rmi0^{+}$ (resp. $0$, $\beta_1^A+\rmi0^{+}$ and $\beta_2^A+\rmi0^{+}$), i.e. the points $\beta_1^A+\rmi0^{+}$ and $\beta_2^A+\rmi0^{+}$ are only branch points when approached from the bottom side of the cut (resp. the top side of the cut), see figure~\ref{analytic_continuation_lambert}. The analytic continuation of the these functions across their cut are summarized in figures \ref{analytic_continuation_lambert} and \ref{ImageW}.

We observe that the only $\Omega_j^A$ which take negative real values are $\Omega_0^A$ and $\Omega_{\pm1}^A$, see figure~\ref{ImageW}. More precisely, $\Omega^{A}_{0}(z)\in\mathbb{R}^{-}$ is equivalent to $z\in[\beta_{1}^{A},0]$, while $\Omega^{A}_{\pm1}(z)\in\mathbb{R}^{-}$ is equivalent to $z\in[\beta_{1}^{A},0]\mp\rmi0^{+}$. This observation is needed below in section~\ref{appendix_wj_yj} in order to determine when the poles $y_j^A$ of the integrands $\chi_A$ and $\eta_A$ from section~\ref{sec_analytic_continuation} are real.

When $A \rightarrow \infty$, the branch point $\beta_2^A$ goes to infinity while $\beta_1^A \rightarrow - \frac{1}{e}$, so that one recovers the analytic structure of the usual Lambert $W_j$. When $A\to0$, the functions $\Omega_j^A(z)$ converge instead to the branches of the logarithm.
\end{subsection}

\begin{subsection}{Functions $w_j^A$ and $y_j^A$} \label{appendix_wj_yj}
We now introduce the functions $w_j^A$ and $y_j^A$ which are used to compute the analytic continuation of the functions $\chi_A$ and $\eta_A$ in section~\ref{sec_analytic_continuation}. The poles of the integrand of the expressions (\ref{def_eta}) and (\ref{def_chi}), are the solutions to the equation
\begin{equation}
\left(4 A+y^2\right) e^{-v-v_0^A+\frac{y^2}{4}} + (A+4) y^2 = 0\;,
\end{equation}
where $v_0^A$ defined in (\ref{def_v0}) is actually equal to $v_0^A=-\log(-\beta_1^A)$. These solutions are given by the generalized Lambert functions as
\begin{equation}
\pm 2\sqrt{-W_j^A\Big(\frac{\rme^{-v-v_0^A}}{4+A}\Big)} = \pm 2 \sqrt{-\Omega_j^A(\rme^{-v-v_0^A})}
\end{equation}
We seek to construct determinations $y_j(v)$ of these solutions which are analytic on the domain $\mathbb{D}^A$ defined in (\ref{D}). We first give determinations $w_j^A(v)$ of the functions $\Omega_j^A(\rme^{-v-v_0^A})$ which belong to the space $\mathcal{F}_{A}$ defined in section~\ref{translation} and coincide with $\Omega_j^A(\rme^{-v-v_0^A})$ when $-\pi<\Im\,v<\pi$. One has
\begin{eqnarray}
\fl
 w^A_0(v)  =
\left\{
	\begin{array}{ll}
		\Omega^{A}_{-[\frac{\Im{v}}{2\pi}]}(\rme^{-v-v_0^A}) & \psi (A) < \Re{v} < 0  \\[3mm]
		\Omega^{A}_0(\rme^{-v-v_0^A}) & \text{otherwise}
	\end{array} \right. \\
\fl
 w^A_j(v)  =
\left\{
	\begin{array}{ll}
		\Omega^{A}_{j-[\frac{\Im{v}}{2\pi}]-\sgn (j)}(\rme^{-v-v_0^A}) & \psi (A) < \Re{v} < 0  \quad \& \quad \sgn (j)\,\Im{v} > 2\pi \left(|j|-\frac{1}{2}\right)\\[3mm]
		\Omega^{A}_{j-[\frac{\Im{v}}{2\pi}]}(\rme^{-v-v_0^A}) & \text{otherwise}
	\end{array} \right. \nonumber
\end{eqnarray}
where $[\,\cdot\,]$ denotes rounding to the nearest integer. Taking the square root, the functions $y_j^A(v)$ analytic on $\mathbb{D}^A$ and giving the poles of $f^A_\eta$ and $f^A_\chi$ are
\begin{eqnarray}
\label{yj}
y^A_0(v)=
\left\{
	\begin{array}{lll}
         (-1)^{-[\frac{\Im v}{2\pi}]} \sqrt{-w_0^A(v)} && \Re{v} > 0 \\
         \sgn(\Im~v - 2\pi j)\sqrt{-w_0^A(v)} && \psi (A) < \Re{v} < 0\\
         \sqrt{-w_0^A(v)} && \Re{v} < \psi (A)
	\end{array}
\right.\;\\
y_j^A(v)=
\left\{
	\begin{array}{lll}
         -\sgn (j)\sqrt{-w_j^A(v)} && \Re{v} > 0\\
         \sgn(\Im~v - 2\pi j)\sqrt{-w_j^A(v)} && \Re{v} < 0\\
	\end{array}
\right.\;,
\end{eqnarray}
From the discussion at the end of section~\ref{section OmegajA}, we observe that $y_{0}^{A}(v)\in\mathbb{R}$ is equivalent to $v\in\mathbb{R}^{+}+2\rmi\pi(n+1/2)$, $n\in\mathbb{Z}$ while $y_{j}^{A}(v)\in\mathbb{R}$ is equivalent for $j\neq0$ to $v\in\mathbb{R}^{+}+2\rmi\pi(j+1/2)$.

We are interested in the analytic continuations of the functions $w_j^A$ and $y_j^A$, which can be computed in terms of the translation operators $\mathcal{T}^{n}_{0,\text{r|l}}$ and $\mathcal{T}^{n}_{\psi,\text{r|l}}$ using (\ref{algebra_AT}). The translation operators act on the functions $w_j^A$ and $y_j^A$ as
\begin{eqnarray}
 \mathcal{T}_{\text{m}}~w_j^A  = w^A_{j-1}\;,\\
 \mathcal{T}_{\text{r}}~w_j^A= \mathcal{T}_{\text{l}}~w_j^A = 
\left\{
\begin{array}{ll}
w^A_{0} & j=0\\
w^A_{-1} & j=1\\
w^A_{j-1} & j\notin\{0,1\}
\end{array}
\right.\;,\\
\mathcal{T}_{\text{m}}~y_j^A  = y^A_{j-1}\;,\\
\mathcal{T}_{\text{r}}~y_j^A = \left\{
\begin{array}{ll}
-y^A_{0} & j=0\\
-y^A_{-1} & j=1\\
y^A_{j-1} & j\notin\{0,1\}
\end{array}
\right.\;,\\
\mathcal{T}_{\text{l}}~y_j^A = \left\{
\begin{array}{ll}
y^A_{0} & j=0\\
y^A_{-1} & j=1\\
y^A_{j-1} & j\notin\{0,1\}
\end{array}
\right.\;.
\end{eqnarray}
Using relations (\ref{algebra_AT}), the analytic continuation of the $y_j^A$ through their cuts is given by
\begin{eqnarray}
 \mathcal{A}^n_{0,\text{l}}~y_j^A =\left\{
	\begin{array}{ll}
		(-1)^n y^A_0 & \text{if} \quad j=n\\
		-y^A_{j+\sgn (n)} & \text{if} \quad j \in \{0\} \cup B_{n-\sgn (n)}\\
		y^A_j & \text{if} \quad j \notin \{0\} \cup B_{n}
	\end{array}
\right.\\
\mathcal{A}_{0,\text{r}}^n y^A_j
=\left\{
	\begin{array}{ll}
		(-1)^n y^A_n \quad & \text{if} \quad j=0\\
		-y^A_{j-\sgn (n)} \quad & \text{if} \quad j \in B_{n}\\
		y^A_j \quad & \text{if} \quad j \notin \{0\} \cup B_{n}
	\end{array}
\right.\;,
\end{eqnarray}
and
\begin{eqnarray}
  \mathcal{A}^n_{\psi,\text{l}}~y_j^A =\left\{
	\begin{array}{ll}
		y^A_n & \text{if} \quad j=0\\
		-y^A_{j-\sgn (n)} & \text{if} \quad j \in  B_{n}\\
		y^A_j & \text{if} \quad j \notin \{0\} \cup B_{n}
	\end{array}
\right.\\
\mathcal{A}_{\psi,\text{r}}^n y^A_j
=\left\{
	\begin{array}{ll}
		y^A_0 \quad & \text{if} \quad j=n\\
		-y^A_{j+\sgn (n)} \quad & \text{if} \quad j \in \{0\} \cup \in B_{n-\sgn (n)}\\
		y^A_j \quad & \text{if} \quad j \notin \{0\} \cup B_{n}
	\end{array}
\right.\;,
\end{eqnarray}
where the sets $B_n$ are defined by (\ref{B}).
\end{subsection}
\end{section}

\begin{section}{Connectedness of the Riemann surface $\mathcal{R}^A$} \label{appendix_riemann}
In this section we prove several results stated in section \ref{sec_analytic_continuation} about the orbits of finite set of integers under translation operators, and its consequences on how the sheets of the Riemann surface $\mathcal{R}^A$ are connected together.

\begin{subsection}{Orbits under the action of operators $T_\text{r}^n$}
We first prove that the orbit of any finite set $P \subset \mathbb{Z}$ under the action of the translation operators $T_\text{r}^n$ contains a unique element $P^*$ which has the same number of positive and negative elements. We also show that the collection of all sets $P \subset \mathbb{Z}$ such that $0 \notin P^*$ is globally stable under the action of $T_\text{r}^n$. 
 
Using the notation $T_\text{r}=T_\text{r}^{-1}$, $T_\text{m}=T_\text{m}^{-1}$, $T_\text{l}=T_\text{l}^{-1}$, let $P$ be a finite set of integers with positive elements $P_{+}=\{k^+_1,\ldots,k^+_h\}$, $0<k^+_1<\ldots<k^+_h$ and negative elements $P_{-}=\{k^-_\ell,\ldots,k^-_1\}$, $k^-_\ell<\ldots<k^-_1<0$, the action of $T_\text{r}$ on $P$ writes:
 \begin{equation}
\label{action_Tr}
\fl\hspace{5mm}
T^{-1}_r P = \left\{\begin{array}{ll}
	\{k^-_\ell + 1,..., k^-_2 +1, k^+_1 +1,...,k^+_h+1\} & \text{if} \quad 0 \in P \; \& \; -1 \in P \\
	\{k^-_\ell + 1,..., k^-_1 +1, 1, k^+_1 +1,...,k^+_h+1\} & \text{if} \quad 0 \in P \; \& \; -1 \notin P \\
	\{k^-_\ell + 1,..., k^-_2 +1, 0, k^+_1 +1,...,k^+_h+1\} & \text{if} \quad 0 \notin P \; \& \; -1 \in P \\
	\{k^-_\ell + 1,..., k^-_1 +1, 0,1, k^+_1 +1,...,k^+_h+1\} & \text{if} \quad 0 \notin P \; \& \; -1 \notin P \\
\end{array}\right.\;.
\end{equation}
In any case, the application of $T^{-1}_r$ decreases the excess number of positive elements in $P$ by exactly $1$ so that for any $P$, the set $P^*=T_r^{|P_-|-|P_+|} P$ is the unique element with $|P_+^*|=|P_-^*|$ in the orbit of $P$ under the action of $T_r$. Let us denote by $O_{P^*}$ the orbit under $T_\text{r}$ of the set $P^*$ and
\begin{eqnarray}
 \mathbb{P}^0 & = \{P \subset \mathbb{Z}, 0 \in P^* \}\;,\\
 \mathbb{P}^* & = \{P \subset \mathbb{Z}, 0 \notin P^* \}\;.
\end{eqnarray}
We want to show that $\mathbb{P}^*$ is stable under the action of the analytic continuation operators $A^n_{\psi | 0, r|l}$. Since the operators $A^n_{\psi | 0, r|l}$ are expressed by (\ref{algebra_AT}) as product of the translation operators $T_\text{r}$, $T_\text{m}$ and $T_\text{l}$, we only need to prove that $\mathbb{P}^*$ is globally stable under $T_\text{m}$ and $T_\text{l}$ (the orbits $O_{P^*}$ being stable under $T_\text{r}$ by construction).

We observe from (\ref{action_Tr}) that $0 \in P^*$ is equivalent to $0\in P$ (respectively $0 \notin P$) if $|P_+|-|P_-|$ is even (resp. odd). Under the replacement $P\to T^{-1}_\text{m}P$, $|P_+|-|P_-|$ is left unchanged if $0\notin P$ and is decreased by one otherwise, which proves the stability under $T_\text{m}$. 

Writing down the action of $T_\text{l}$ on a set $P \in \mathbb{P}^*$, wee see that in every case, under the replacement $P\to T^{-1}_\text{m}P$, $|P_+|-|P_-|$ is left unchanged or decreased by two, which concludes the proof. The collections of sets $\mathbb{P}^0$ and $\mathbb{P}^*$ are thus stable under the action of analytic continuation operators so that the associated Riemann surfaces $\mathcal{R}^A_0$ and $\mathcal{R}^A_*$ form two disjoint components of $\mathcal{R}^A$.
\end{subsection}

\begin{figure}
\begin{center}
 \includegraphics[scale=0.7]{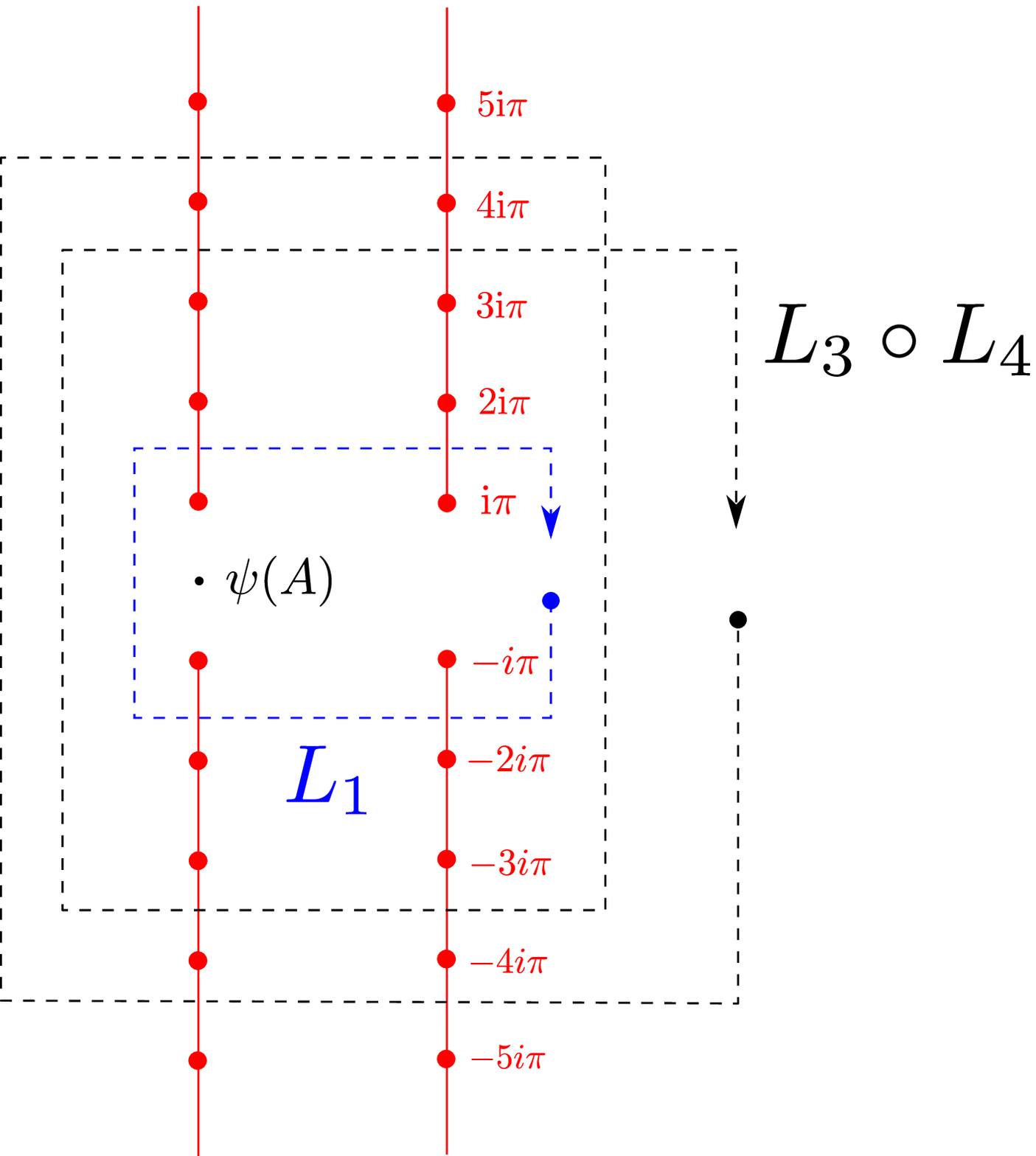}
 \caption{The paths corresponding to the application of the operator $L_1$ (blue dashed line) and to the operators $L_3 \circ L_4$ (black dashed line) projected on the complex plane.}
 \label{paths_riemann}
 \end{center}
 \end{figure}

\begin{subsection}{Connectedness of $\mathcal{R}_*^A$}
In this section we prove that the Riemann surface $\mathcal{R}_*^A$ representing actual eigenstates is connected for $0 \leq A < \infty$, while for $A \rightarrow \infty$ it is split into an infinity of disjoint connected components. Let us introduce for $k\in\mathbb{N}$
\begin{equation}
C_k = \left\{ P \subset \mathbb{Z}, \bigg| | P^*_\text{odd} | - | P^*_\text{even} | \bigg| = 2k \right\}
\end{equation}
and $\mathcal{C}_k$ the collection of sheets in $\mathcal{R}^A$ indexed by the sets $P$ in $C_k$. We first show that the sets $C_k$ are stable and consist of a single orbit under the action of operators $A^n_{0,\text{l}}$ and $A^n_{0,\text{r}}$. Then, we show that any set in $\mathbb{P}^*$ can be constructed from the empty set by repeated applications of operators $A^n_{\psi | 0, r|l}$.

One easily checks that the action of operators $T_\text{r}$ and $T_\text{l}$ on any set $P$ does not change the excess number of elements of a given parity $\bigg| | P^*_\text{odd} | - | P^*_\text{even} | \bigg|$ in $P^*$, so the sets $C_k$ are stable under these actions by construction. It follows from (\ref{algebra_AT}) that they are then stable under the action of $A^n_{0,\text{l}}$ and $A^n_{0,\text{r}}$. Let
\begin{equation}
 P_k = \{2j-1, 1 \leq j \leq 2k \} \cup \{-2j+1, 1 \leq j \leq 2k \}\;.
\end{equation}
$P_k$ is an element of $C_k$ which can be constructed by repeated applications of operators $A^n_{0,\text{l}}$ and $A^n_{0,\text{r}}$ on any set $P \in C_k$. This is proved by remarking that the action of $A^n_{0,\text{l}}$ on $P \ni n$ removes the element $n$ if $n-1 \in P$ while the action of $A^n_{0,\text{r}}$ removes $n$ from $P$ if $0 \in P$ (respectively $0 \notin P$) if $n$ is odd (respectively $n$ is even). Hence, starting from $P^*$ in the orbit $O_P$, one can reach $P_k$ by successive substitutions $P \rightarrow A^n_{0,\text{l} | \text{r}}$, so that the components $\mathcal{C}_k$ are connected through the cuts at $\Re v = 0$.

To show that $\mathcal{R}^A_*$ is connected for $0 \leq A < \infty$, let us define the operator $L_n =  A^n_{0,\text{l}} A^n_{\psi,\text{l}}A^{-n}_{\psi,\text{r}} A^{-n}_{0,\text{r}}$ (the corresponding paths are represented in figure \ref{paths_riemann}). The set $P_k$ is obtained from the empty set as
\begin{equation}
 P_k = L_1 \circ L_2 \circ ... \circ L_n~\emptyset
\end{equation}
so that all component $\mathcal{C}_k$, $k\in\mathbb{N}$ are connected to one another. In the limit $A \rightarrow \infty$, the paths through the cuts at $\Re v = \psi(A)$ do not exist anymore, and the $\mathcal{C}_k$ then become independent connected components of $\mathcal{R}_*^A$.
\end{subsection}
\end{section}

\end{document}